\documentclass{raa}            % referee version: for submission

\usepackage{graphicx,times}             %for PS/EPS graphics inclusion, new

\begin{document}

   \title{Preliminary results of CCD observations of Himalia at Yunnan Observatories in 2015
%\,$^*$
%\footnotetext{$*$ Supported by the National Natural Science Foundation of China.}
}
%   \subtitle{I. Place Your Subtitle Here}

   \volnopage{Vol.0 (201x) No.0, 000--000}      %%preserved for Editor. DOn't remove!
   \setcounter{page}{1}          %%starting page, preserved for Editor. DOn't remove!

   \author{Huan-Wen Peng
      \inst{1,2,3}
   \and Na Wang
      \inst{1}
   \and Qing-Yu Peng
      \inst{1}
   }
%% Here is an example of three authors come from different institutes.
%% For single author or all the authors from an institute, use "\inst{}" only

   \institute{Sino-French Joint Lavoratory for Astrometry, Dynamics and Space Science, Jinan University, Guangzhou 510632, China;{\it ~tpengqy@jnu.edu.cn}\\
        \and
            Yunnan Observatories, Chinese Academy of Sciences, Kunming 650216, China \\
        \and
            University of Chinese Academy of Sciences, Beijing 100049, China\\}

   \date{Received~~2016 6 20; accepted~~2016~~8 26}

\abstract{In order to study the potential in high precision CCD astrometry of irregular satellites, we have made experimental observations for Himalia, the sixth and irregular satellite of Jupiter. A total of 185 CCD observations were obtained by using the 2.4 m telescope and 1 m telescope at Yunnan Observatories over ten nights. Preliminary analysis for the observations were made, including the geometric distortion, the atmospheric refraction, and also the phase effect. All positions of Himalia are measured relative to the reference stars from the catalogue UCAC4 in each CCD field of view. The theoretical positions of Himalia were retrieved from the Institute de M\'{e}chanique C\'{e}leste et de Calcul des \'{E}ph\'{e}m\'{e}rides, while the positions of Jupiter were obtained based on the planetary theory INPOP13C. The results show that the mean (O-C) (observed minus computed) residuals are -0.004 and -0.002 arcsec in right ascension and declination, respectively. The standard deviation of (O-C) residuals are estimated at about 0.04 arcsec in each direction.
\keywords{astrometry --- planets and satellites: individual: Himalia --- methods: observational}
}

   \authorrunning{Huan-Wen Peng, Na Wang \& Qing-Yu Peng}            %author_head in even pages
   \titlerunning{Preliminary results of CCD observations of Himalia}  % title_head in odd pages

   \maketitle
%% The author head (on even pages) and the title head (on odd pages) will be
%% automatically extracted from \author{} and \title{}. Whenever the title is too long,
%% you will be asked to supply a shorter one by inserting either \authorrunning{} or
%% \titlerunning{} before \maketitle. Anyway, you can specify your own heads.
%%
%%
%% Note: In the following text body of your manuscript, please note several differences from
%%       other major journals:
%% (1) \subsection{Please Capitalize the First Letter of Each Notional Word in Subsection Title}
%% (2) Please Capitalize the First Letter of Each Notional Word in all tables' captions

%
%________________________________________________ sections below
%
\section{Introduction}           %% first-level sections will be auto-capitalized
\label{sect:intro}

The irregular satellites are natural moons of the giant planets in our solar system. Unlike the regular ones, these irregular satellites are smaller, having more distant, highly eccentric and highly inclined orbits (Nicholson et al.~\cite{nicholson2008}; Grav et al.~\cite{grav2015}). It is widely accepted that these objects are closely related to early solar system formation, because they are believed to have been heliocentric asteroids before being captured by a giant planet's gravity (Colombo \& Franklin~\cite{colombo1971}; Heppenheimer \& Porco~\cite{heppenheimer1977}; Pollack et al.~\cite{pollack1979}; Sheppard \& Jewitt~\cite{sheppard2003}; Agnor \& Hamilton~\cite{agnor2006}; Nesvorn\'{y} et al.~\cite{nesvorny2007},~\cite{nesvorny2014}). In order to further understand their dynamics and physics, routine observations both from the Earth and from the spacecrafts are needed (Jacobson et al.~\cite{jacobson2012}). In comparison with the regular ones, the ephemerides of the irregular satellites have relatively worse precision. Thus lots of high precision astrometric observations are required to improve their ephemerides.

Himalia is the largest member of the Jovian outer irregular satellites (Grav et al.~\cite{grav2015}). It has been discovered by Perrine at Lick Observatory in 1904 (Perrine~\cite{perrine1905}). Continuous observations were made after its discovery, including the first computation of diameter, 170$\pm$20km, by Cruikshank (\cite{cruikshank1977}). While the high precision space observations were made after Himalia was first visited in a flyby by the Cassini space probe in 2000 (Porco et al.~\cite{porco2003}), long-term CCD observations of ground-based telescopes have been obtained and also show great potential in deriving high precision positions for the irregular satellites (Gomes-J\'{u}nior et al.~\cite{gomes2015}). In our recent research work of Phoebe, the ninth and irregular satellite of Saturn, positional precision of 0.$''$04 in each direction were obtained (Peng et al.~\cite{peng2015}). In order to obtain high precision astrometric results, a series of error effects should be taken into account, especially for the geometric distortion (called GD hereafter). Previous research works (Peng et al.~\cite{peng2012}; Zhang, Peng \& Zhu~\cite{zhang2012}) had shown obvious GD effects in the CCD field of view of the 2.4 m telescope and the 1 m telescope at Yunnan Observatories. More recent research works (Peng et al.~\cite{peng2015}; Wang et al.~\cite{wang2015}) again confirmed the GD effects.

The contents of this paper are arranged as follows. In Section 2, the astrometric observations are described. Section 3 presents the reduction details of the observations and Section 4 shows the results. In Section 5, we make the discussions. Finally, in Section 6, conclusions are drawn.

%% Authors can give a citation as 'Michel et al. 1992'.
%% You may also use \cite, \citep and \citet for citation, and use Table~1 or Figure~1
%% and so forth. Using \ref and \label for cross-references of Tables/Figures
%% is a good way in adjusting/adding/removing text, tables or figures.

\section{Astrometric Observations}
\label{sect:observation}

Ten nights of astrometric observations of Himalia were carried out by two
telescopes at Yunnan Observatories in 2015. Specifically, eight nights of CCD observations were made by the Yunnan Faint Object Spectrograph and Camera (YFOSC) instrument attached to the 2.4 m telescope at Lijiang Observatory (longitude E 100$^\circ$1$'$51$''$, latitude N 26$^\circ$42$'$32$''$, height 3193 m above sea level),
and two nights of CCD observations were made by the 1 m telescope at Yunnan Observatories (longitude E 102$^\circ$47$'$18$''$, latitude N 25$^\circ$1$'$46$''$, height 2000 m above sea level). The detailed specifications of the two telescopes and CCD detectors are listed in Table 1. The observational dates were chosen according to the epoch when Jupiter at it's opposition which is February 6, 2015. However, no observations were made during February 1 to February 6, because the Jovian system and the Moon were very close.

\begin{table}[!b]
\begin{center}
\caption[]{Specifications of the 2.4 m telescope and 1 m telescope at Yunnan Observatories and the corresponding CCD detectors. Column 1 shows the parameters and the following columns list their values for the two telescopes.}
  \label{Tab1}
  \begin{tabular}{rrr}
  \hline\noalign{\smallskip}
  Parameters                           & 2.4 m telescope          & 1 m telescope            \\
  \hline\noalign{\smallskip}
  Approximate focal length             & 1920cm                   & 1330cm                   \\
  F-Ratio                              & 8                        & 13                       \\
  Diameter of primary mirror           & 242cm                    & 102cm                    \\
  CCD field of view(effective)         & 9$'\times$9$'$           & 7$'\times$7$'$           \\
  Size of CCD array(effective)         & 1900$\times$1900         & 2048$\times$2048         \\
  Size of pixel                        & 13.5$\mu\times$13.5$\mu$ & 13.5$\mu\times$13.5$\mu$ \\
  Approximate angular extent per pixel & 0.$''$286$/$pixel        & 0.$''$209$/$pixel        \\
  \noalign{\smallskip}\hline
\end{tabular}
\end{center}
\end{table}

\begin{table}[!b]
\begin{center}
\caption[]{CCD observations for Himalia and calibration fields by using the 2.4 m telescope and 1 m telescope at Yunnan Observatories. Column 1 shows the observational dates. Column 2 lists the dense star fields observed. Column 3 and Column 4 list the numbers of observations and filter for dense star fields and Himalia, respectively. Column 5 shows which telescope was used.}
  \label{Tab2}
  \begin{tabular}{ccccc}
  \hline\noalign{\smallskip}
  Observation Dates& Calibration fields &                & Himalia        & Telescope \\
  \hline\noalign{\smallskip}
                   & Dense Star Fields  & No. and filter & No. and filter &           \\
  \hline\noalign{\smallskip}
  2015-01-31       & NGC1664            & 44I            & 21I            & 2.4 m     \\
  2015-02-07       & NGC2324            & 44I            & 25I            & 2.4 m     \\
  2015-02-08       & NGC2324            & 44I            & 14I            & 2.4 m     \\
  2015-02-09       & NGC2324            & 44I            & 18I            & 2.4 m     \\
  2015-02-10       & NGC1664            & 44I            & 18I            & 2.4 m     \\
  2015-02-11       &                    &                & 18I            & 2.4 m     \\
  2015-02-12       &                    &                & 18I            & 2.4 m     \\
  2015-02-13       &                    &                & 19I            & 2.4 m     \\
  2015-02-12       & M35                & 60I            & 14I            & 1 m       \\
  2015-02-14       &                    &                & 20I            & 1 m       \\
  \hline\noalign{\smallskip}
  Total            &                    & 280I           & 185I           &           \\
  \noalign{\smallskip}\hline
\end{tabular}
\end{center}
\end{table}

A total of 185 CCD frames for Himalia have been obtained, as well as 280 CCD frames of calibration fields. The details of the distribution of observations with respect to the observational dates for the two telescopes are listed in Table 2. It can be seen that 151 CCD frames for Himalia were obtained from the 2.4 m telescope, and 34 CCD frames for Himalia were obtained from the 1 m telescope. The exposure time for each CCD frame is ranged from 18 seconds to 120 seconds, depending on the diameter of primary mirror and meteorological conditions. Calibration fields were observed following with Himalia, except for 4 nights that the observations were subjected to rapidly changing weather conditions.

\section{Reduction of the observations}
\label{sect:reduction}

After the preprocessing steps including bias and flat correction were performed,  image centering was applied by using the two-dimensional Gaussian fit algorithm (Li, Peng \& Han~\cite{li2009}). The same procedure were used to process the CCD frames of calibration fields which are open clusters, and then the GD patterns were derived. More details about the deriving of the GD pattern are presented in Peng et al. (\cite{peng2012}). Fig.1 shows the GD patterns for the 2.4 m telescope and the 1 m telescope.

\begin{figure}[htb]
\centering
\includegraphics[width=14cm,angle=0]{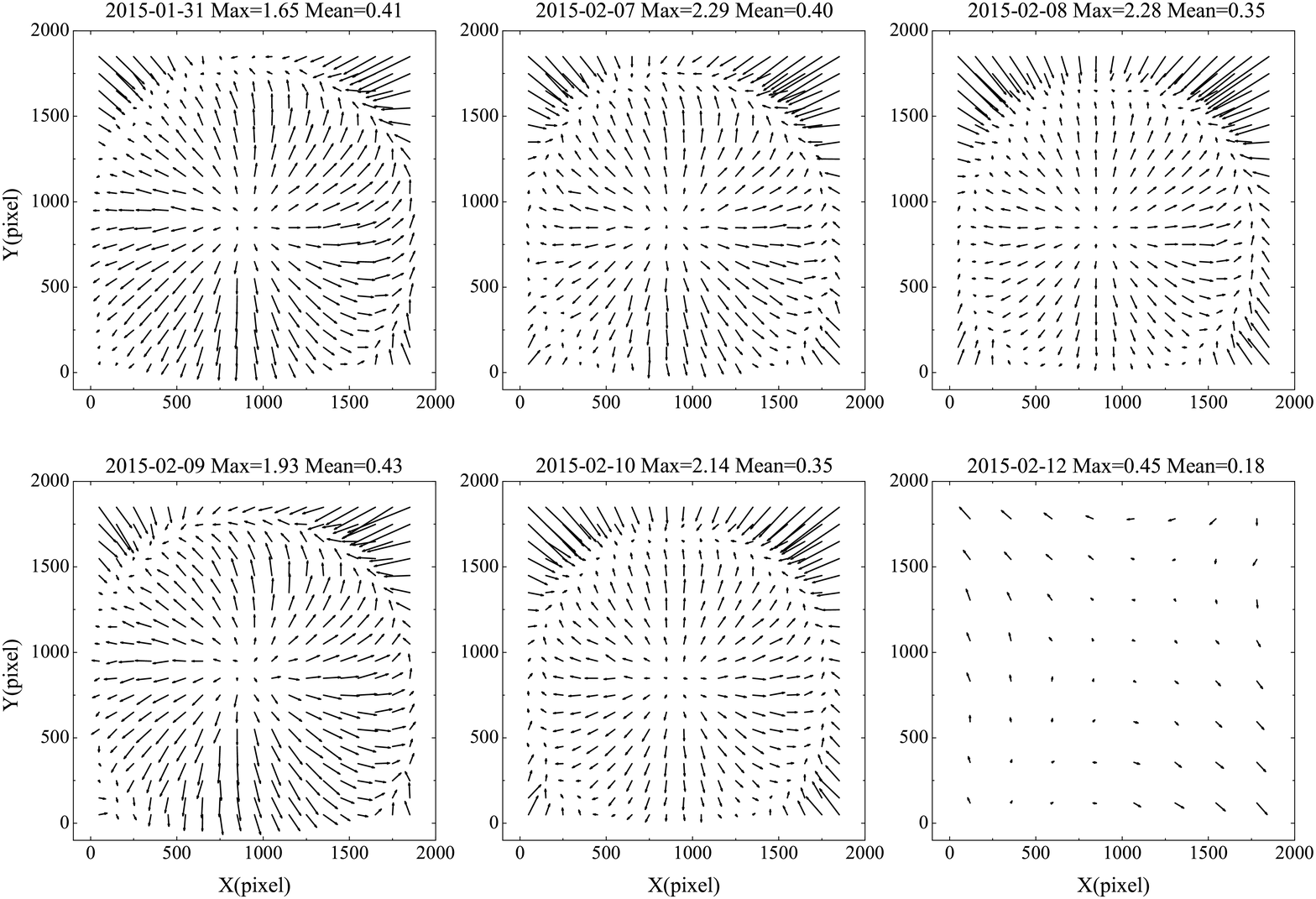}
\caption{GD patterns for the 2.4 m telescope and 1 m telescope of Yunnan Observatories.
The former five panels show the GD patterns derived from the CCD observations of NGC1664 and NGC2324 with the 2.4 m telescope. The rightmost panel in second row shows the GD pattern derived from the CCD observations of M35 with the 1 m telescope. All observations were made with a filter-I. In each panel, the observational dates, the maximum GD values and the mean GD values are listed on the top of all panels in units of pixels. A factor of 200 is used to exaggerate the magnitude of each GD vector.}
\label{Fig1}
\end{figure}

As mentioned above, the calibration fields were not observed on every night. GD corrections were applied on each night if the GD pattern is available, otherwise the GD pattern of the nearest night is used. Further more, we have made several experiments for the CCD frames of Himalia obtained with the 2.4 m telescope. GD corrections were applied by using only one of the GD patterns in each of the experiments. The results shown that the absolute difference in positional measurement made when using a GD pattern from another night is within 0.$''$005. More details are presented in our previous research work (Wang et al.~\cite{wang2015}), in which the influence on (O-C) residuals by using different GD schemes was studied.

The catalogue UCAC4 (Zacharias et al.~\cite{zacharias2013}) was chosen to match the reference stars in all CCD frames. The minimum and maximum numbers of UCAC4 reference stars available for Himalia astrometric reduction for the observations made with the 2.4 m telescope are 7 and 19, respectively. For the observations made with the 1 m telescope, these two numbers are 5 and 9, respectively. Observed positions of Himalia were derived relative to these UCAC4 reference stars by using a four-constant plate model. However, this is accurate only after all the astrometric effects, including GD effects, are taken into account (Peng et al.~\cite{peng2012}).

According to the illustration presented in Lindegren (\cite{lindegren1977}), the phase effect have direct influence on the positional measurements of planets and natural satellites in our solar system. Phase corrections should be concerned and applied for these objects. By using the equation (14) presented in Lindegren (\cite{lindegren1977}), the phase effect of Himalia were calculated. Table 3 shows the details. It can be seen that the maximum value of phase effect according to the observation time is as small as 0.$''$0005. This means the phase effect of Himalia for our observations are negligible.

\begin{table}[htb]
\begin{center}
\caption[]{Phase effect for Himalia with respect to the observation time. Column 1 shows several typical observation time. Column 2 lists the distance of Himalia from the Earth. Column 3 and Column 4 list the apparent diameter and the phase angle, respectively. Column 5 shows the extincted V magnitude. Column 6 lists the phase effect for Himalia.}
  \label{Tab3}
  \begin{tabular}{cccccc}
  \hline\noalign{\smallskip}
  Observation Time         & Distance & Apparent Diameter & Phase Angle          & V    & Phase Effect \\
  \hline\noalign{\smallskip}
  (UTC)                    & (AU)     &                   &                      & (extincted)&\\
  \hline\noalign{\smallskip}
  2015-01-31 15$^h$ 13$^m$ &4.418     &0.$''$1            &+1$^\circ$19$'$26$''$ &15.80 &0.$''$0004 \\
  2015-02-13 16$^h$ 01$^m$ &4.416     &0.$''$1            &+1$^\circ$23$'$12$''$ &15.77 &0.$''$0005 \\
  2015-02-12 15$^h$ 10$^m$ &4.415     &0.$''$1            &+1$^\circ$10$'$50$''$ &15.78 &0.$''$0004 \\
  2015-02-14 15$^h$ 53$^m$ &4.418     &0.$''$1            &+1$^\circ$35$'$06$''$ &15.77 &0.$''$0005 \\
  \noalign{\smallskip}\hline
\end{tabular}
\end{center}
\end{table}

The IAU-SOFA (Standards of Fundamental Astronomy) library (Wallace~\cite{wallace1996}) was used to calculate the topocentric apparent positions of the reference stars in all CCD frames. The IAU 2006/2000A precession-nutation models (Capitaine \& Wallace~\cite{capitaine2006}) are used in the calculations. In order to derive the GD patterns accurately, the atmospheric refraction effect (the standard model should be precise enough (Peng et al.~\cite{peng2012})) is added to the positional reduction of reference stars.

\section{Results}
\label{sect:results}

The observed positions of Himalia were compared to the ephemerides retrieved from Institute de M\'{e}chanique C\'{e}leste et de Calcul des \'{E}ph\'{e}m\'{e}rides which includes satellites theory by Emelyanov (\cite{emelyanov2005}), and the theoretical positions of Jupiter were obtained by using planetary theory INPOP13C (Fienga et al.~\cite{fienga2015}). For comparison, the ephemerides computed by Jet Propulsion Laboratory Horizons ephemeris service (Giorgini et al.~\cite{giorgini1996}) were also obtained, including the satellites theory JUP300 (Jacobson~\cite{jacobson2013}) and the planetary theory DE431 (Folkner et al.~\cite{folkner2014}).

\begin{figure}[htb]
\centering
\includegraphics[width=14cm,angle=0]{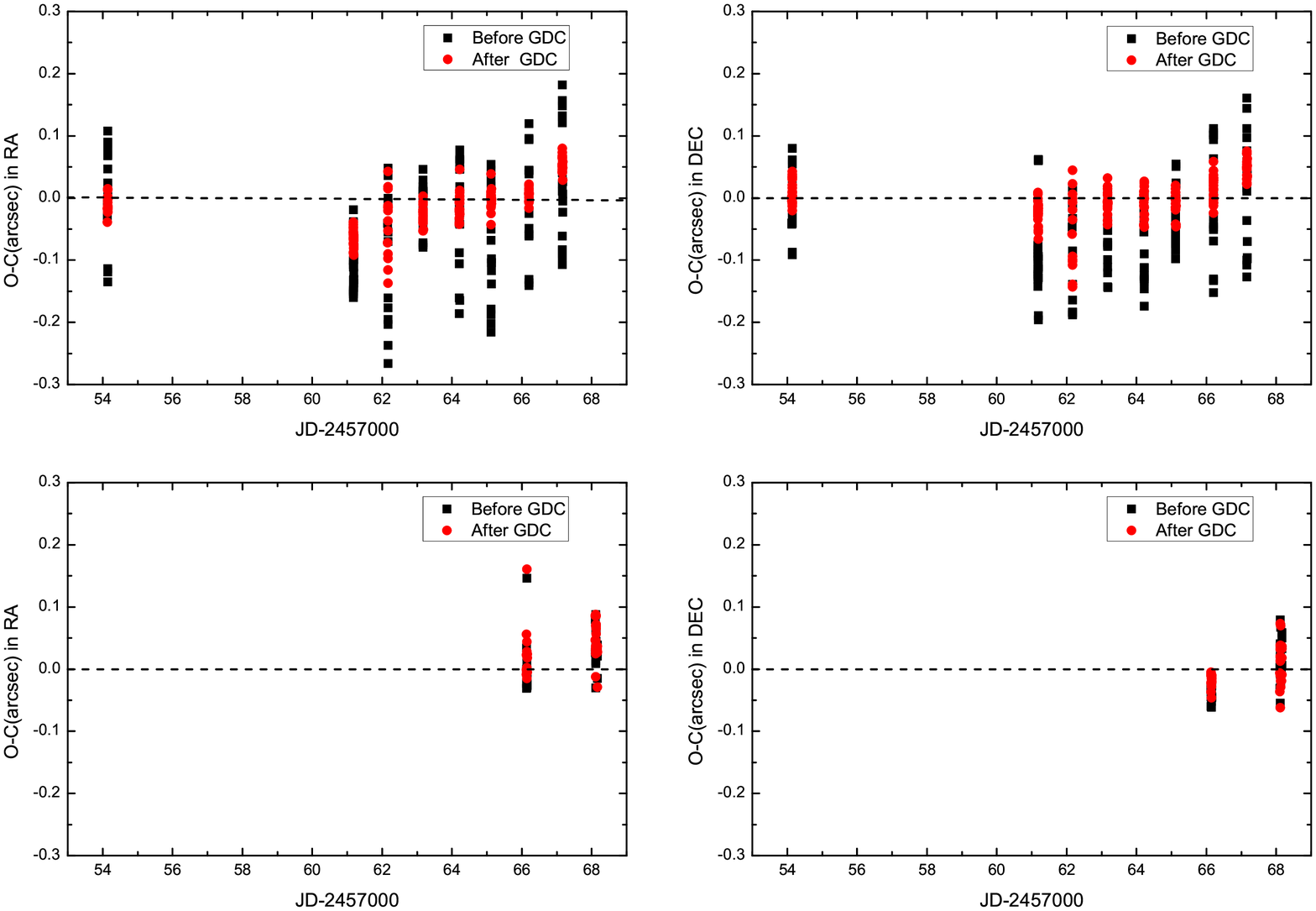}
\caption{(O-C) residuals of the topocentric apparent positions of Himalia compared to the satellites theory by Emelyanov (\cite{emelyanov2005}), together with the planetary theory INPOP13C, with respect to the Julian Dates. The upper two panels show the (O-C) residuals for the CCD observations of the 2.4 m telescope, and the bottom two panels show the (O-C) residuals for the CCD observations of the 1 m telescope. The dark points represent the (O-C) residuals before GD corrections and the red ones represent the (O-C) residuals after GD corrections.}
\label{Fig2}
\end{figure}

\begin{table}[htb]
\begin{center}
\caption[]{Statistics of (O-C) residuals of the positions of Himalia before and after GD corrections. Column 1 shows the observational dates and the telescope. The second column shows whether the GD corrections are made. The following columns list the mean (O-C) residuals and its standard deviation (SD) in right ascension and declination, respectively. All units are in arcseconds.}
  \label{Tab4}
  \begin{tabular}{cccccc}
  \hline\noalign{\smallskip}
  Observation Date & GDC    &$<O-C>$ & SD     &$<O-C>$ & SD        \\
  and Telescope    &        & RA     &        & DEC    &           \\
  \hline\noalign{\smallskip}
  2015-01-31       & Before &  0.021 & 0.072  & -0.007 & 0.049     \\
  2.4 m            & After  & -0.009 & 0.013  &  0.009 & 0.018     \\
  \hline\noalign{\smallskip}
  2015-02-07       & Before & -0.106 & 0.041  & -0.092 & 0.065     \\
  2.4 m            & After  & -0.065 & 0.014  & -0.023 & 0.018     \\
  \hline\noalign{\smallskip}
  2015-02-08       & Before & -0.096 & 0.107  & -0.087 & 0.083     \\
  2.4 m            & After  & -0.042 & 0.054  & -0.052 & 0.062     \\
  \hline\noalign{\smallskip}
  2015-02-09       & Before & -0.011 & 0.037  & -0.062 & 0.051     \\
  2.4 m            & After  & -0.029 & 0.016  & -0.006 & 0.018     \\
  \hline\noalign{\smallskip}
  2015-02-10       & Before & -0.028 & 0.084  & -0.083 & 0.054     \\
  2.4 m            & After  & -0.011 & 0.021  & -0.009 & 0.024     \\
  \hline\noalign{\smallskip}
  2015-02-11       & Before & -0.072 & 0.090  & -0.035 & 0.046     \\
  2.4 m            & After  & -0.002 & 0.017  & -0.009 & 0.019     \\
  \hline\noalign{\smallskip}
  2015-02-12       & Before & -0.007 & 0.078  & -0.002 & 0.087     \\
  2.4 m            & After  &  0.005 & 0.010  &  0.018 & 0.022     \\
  \hline\noalign{\smallskip}
  2015-02-13       & Before &  0.023 & 0.097  &  0.009 & 0.089     \\
  2.4 m            & After  &  0.054 & 0.012  &  0.046 & 0.017     \\
  \hline\noalign{\smallskip}
  2015-02-12       & Before &  0.002 & 0.047  & -0.042 & 0.011     \\
  1 m              & After  &  0.022 & 0.045  & -0.019 & 0.012     \\
  \hline\noalign{\smallskip}
  2015-02-14       & Before &  0.034 & 0.029  &  0.017 & 0.035     \\
  1 m              & After  &  0.045 & 0.031  &  0.009 & 0.035     \\
  \hline\noalign{\smallskip}
  Total            & Before & -0.025 & 0.085  & -0.038 & 0.072     \\
                   & After  & -0.004 & 0.044  & -0.002 & 0.036     \\
  \noalign{\smallskip}\hline
\end{tabular}
\end{center}
\end{table}

Fig.2 shows the (O-C) residuals of the positions for Himalia with respect to the observational epochs. Table 4 lists the statistics of (O-C) residuals of Himalia before and after GD corrections. It can be seen that the internal agreement or precision for individual night has significantly improved after GD corrections for the 2.4 m telescope, while slightly changed for the 1 m telescope due to the GD effect are quite smaller than the 2.4 m telescope, as showed in Fig.1. The mean (O-C) residuals for all data sets after GD corrections are -0.$''$004 and -0.$''$002 in right ascension and declination, respectively. And its standard deviation are 0.$''$044 and 0.$''$036 in each direction.

\section{Discussions}
\label{sect:discussion}

In order to check the precision of our CCD observations for Himalia, two different ephemeris were selected to calculate theoretical positions, and then the comparisons were made with the observed ones. The first ephemeris developed by the IMCCE includes the satellites theory by Emelyanov (\cite{emelyanov2005}) and the planetary theory INPOP13C (Fienga et al.~\cite{fienga2015}). The second ephemeris which includes the satellites theory JUP300 (Jacobson~\cite{jacobson2013}) and the planetary theory DE431 (Folkner et al.~\cite{folkner2014}) is developed by the JPL.

\begin{figure}[htb]
\centering
\includegraphics[width=14cm,angle=0]{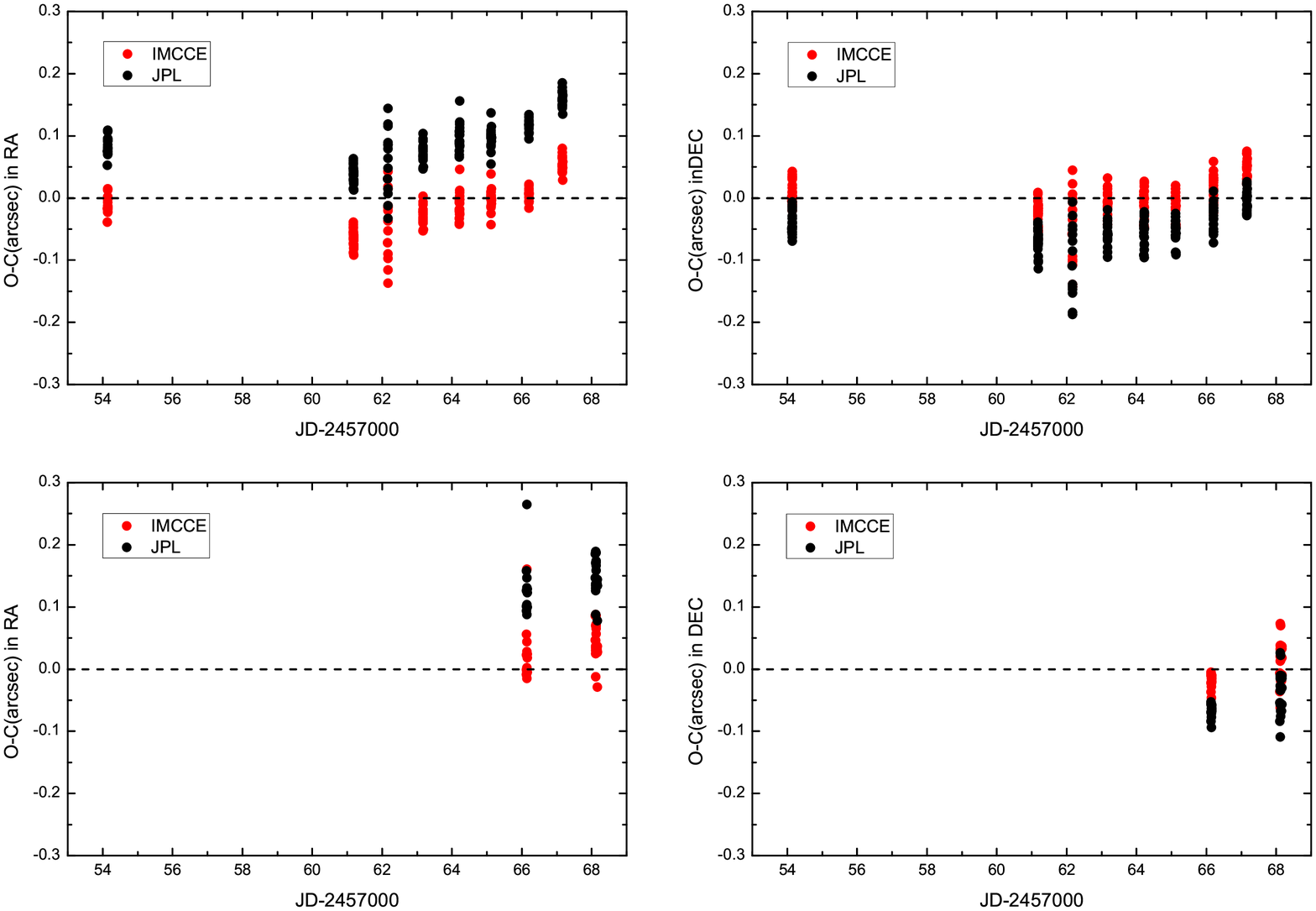}
\caption{(O-C) residuals of the topocentric apparent positions of Himalia compared to two ephemerides. The red points represent the (O-C) residuals after GD corrections for the ephemerides developed by the IMCCE which includes satellites theory by Emelyanov (\cite{emelyanov2005}) and planetary theory INPOP13C. The dark points represent the (O-C) residuals after GD corrections for the ephemerides developed by the JPL which includes satellites theory JUP300 and planetary theory DE431. The upper two panels show the (O-C) residuals for the CCD observations of 2.4 m telescope, and the bottom two panels show the (O-C) residuals for the CCD observations of 1 m telescope. }
\label{Fig3}
\end{figure}

\begin{table}[htb]
\begin{center}
\caption[]{Statistics of (O-C) residuals for Himalia compared to different ephemeris. Column 1 shows the number of CCD observations. Column 2 shows the different ephemeris used. The following columns list the mean (O-C) residuals and its standard deviation in right ascension and declination, respectively. All units are in arcseconds.}
\label{Tab5}
  \begin{tabular}{cccccc}
  \hline\noalign{\smallskip}
  N   &  Ephemerides                  & $<O-C>$ & SD    & $<O-C>$ & SD    \\
      &                               & RA      &       & DEC     &       \\
  \hline\noalign{\smallskip}
  185 & Emelyanov (2005)/INPOP13C     & -0.004  & 0.044 & -0.002  & 0.036 \\
      & JUP300/DE431                  &  0.098  & 0.045 & -0.051  & 0.035 \\
  \noalign{\smallskip}\hline
\end{tabular}
\end{center}
\end{table}

\begin{table}[!htbp]
\begin{center}
\caption[]{Comparisons made with other observations retrieved from the MPC. Column 1 shows the IAU code of observatories. Column 2 lists the number of CCD observations. The following columns show the mean (O-C) residuals and its standard deviation (SD) in right ascension and declination, respectively. All positions of Himalia are observed topocentric astrometric positions. All units are in arcseconds.}
\label{Tab6}
  \begin{tabular}{ccccccc}
  \hline\noalign{\smallskip}
  Observatory    &  Frame & $<O-C>$ & SD    & $<O-C>$ & SD    & Time(yr)  \\
      code       &  no.   & RA      &       & DEC     &       &           \\
  \hline\noalign{\smallskip}
  689            & 68     & -0.046  & 0.149 & -0.006  & 0.131 & 2000-2001  \\
  689            & 106    &  0.005  & 0.148 & -0.020  & 0.143 & 2001-2003  \\
  689            & 103    & -0.001  & 0.182 & -0.011  & 0.191 & 2003-2007  \\
  415            & 24     & -0.007  & 0.106 &  0.071  & 0.088 & 2008       \\
  809            & 23     & -0.048  & 0.092 &  0.019  & 0.047 & 2007-2009  \\
  511            & 357    & -0.018  & 0.049 & -0.011  & 0.061 & 1997-2008  \\
  874            & 238    & -0.075  & 0.175 & -0.007  & 0.034 & 1992-2014  \\
  874            & 560    &  0.005  & 0.069 & -0.018  & 0.053 & 1992-2014  \\
  874            & 56     & -0.036  & 0.113 & -0.019  & 0.069 & 1992-2014  \\
  This paper     & 185    & -0.004  & 0.044 & -0.002  & 0.036 & 2015       \\
  \noalign{\smallskip}\hline
\end{tabular}
\end{center}
\end{table}

Fig.3 shows the (O-C) residuals of the topocentric apparent positions of Himalia in comparison with different ephemerides. It can be seen an obvious systematic deviation between the two ephemerides. While the dispersion for each night between the two ephemerides is similar.
Table 5 shows the statistics of (O-C) residuals of Himalia after GD corrections for different ephemeris. The mean of (O-C) residuals after GD corrections for ephemerides calculated by IMCCE are -0.$''$004 and -0.$''$002 in right ascension and declination, respectively. The values of standard deviation in each direction are 0.$''$044 and 0.$''$036. For ephemerides computed by JPL, the mean (O-C) residuals after GD corrections are 0.$''$098 and -0.$''$051 in right ascension and declination, respectively. The values of standard deviation in each direction are 0.$''$045 and 0.$''$035. A good agreement can be seen for precision between both ephemeris. However, it can be found quite differences for the mean (O-C) residuals in right ascension and declination between two ephemeris.

To compare our CCD observations with previous ones, some major observational statistics of Himalia retrieved from the Minor Planet Center (MPC) are listed. The ephemeris was developed by the IMCCE which includes the satellites theory by Emelyanov (\cite{emelyanov2005}) and the planetary theory INPOP13C (Fienga et al.~\cite{fienga2015}). Table 6 shows the comparisons of the mean (O-C) residuals and its standard deviation. The positions of Himalia are observed topocentric astrometric positions. It can be seen that our observations have comparable precision.

In Fig.2 and Fig.3, we can see that the dispersion of positions of Himalia with 2.4 m telescope at the third night is somewhat worse than other nights after GD corrections. We give the explanation that the serious meteorological conditions were happened on that night, and this would influence the positional measurements. Also it can be seen a systematic trend appearing in the O-Cs in Fig.3, especially in right ascension. This may be caused by the existence of zonal errors in the star catalogue. Another reason might come from the ephemeris. More observations are needed to confirm this in future.

\section{Conclusions}
\label{sect:conclusion}

In this paper, we have made a preliminary analysis of CCD astrometric observations for Himalia. a total of 185 CCD observations were taken with the 2.4 m telescope and 1 m telescope at Yunnan Observatories. A series of error effects were analyzed, including the geometric distortion, the atmospheric refraction, and also the phase effect. The positional precision for Himalia has significantly improved after GD corrections. Comparisons with two different ephemerides have been made. The results have shown that the mean (O-C) residuals of Himalia are -0.$''$004 and -0.$''$002 by using ephemerides developed by the IMCCE in right ascension and declination, respectively. And the standard deviation are 0.$''$044 and 0.$''$036 in each direction.

In consideration of the fewer number of reference stars in each CCD frame of Himalia, the high-order plate model can't be applied. In order to use the low-order plate model which is the four-constant plate model in this paper, the GD effects should be corrected accurately. Further more, the precision of star catalogue also has direct influence on positional measurements of targets. In near future, a new catalogue developed by the ESA astrometry satellite Gaia (Lindegren et al.~\cite{lindegren2008}) will be released soon since the Gaia probe has been launched on December 19, 2013. The precise star positions to be derived by the new catalogue will render better predictions with the only source of error being the ephemerides (de Bruijne~\cite{bruijne2012}; Gomes-J\'{u}nior et al.~\cite{gomes2015}). We believed that the higher astrometric precision of positions for irregular satellites will be achieved.

\begin{acknowledgements}
We acknowledge the support of the staff of the Lijiang 2.4 m telescope. Funding for the telescope has been provided by CAS and the People¡¯s Government of Yunnan Province. Also we acknowledge the support of the staff of the 1 m telescope at Yunnan Observatories. This research work is financially supported by the National Natural Science Foundation of China (grant nos. U1431227,11273014).
\end{acknowledgements}

%\appendix                  %%appendicial material is supported
%
%\section{This shows the use of appendix}
%A postscript file is actually an ASCII text file (you may even edit it).
%However, you need to transfer a PDF file or any compressed or packaged
%file in binary mode when using FTP.
%
%\section{What is SCI?}
%SCI is the abbreviation of Science Citation Index system powered by
%the Institute for Scientific Information (ISI). For details please
%visit {\it http://apps.isiknowledge.com}.

\label{lastpage}

\end{document}